       \newcommand{\beq}{\begin{equation}}
       \newcommand{\eeq}{\end{equation}}
       \newcommand{\beqa}{\begin{eqnarray}}
       \newcommand{\eeqa}{\end{eqnarray}}
       \newcommand{\beqas}{\begin{eqnarray*}}
       \newcommand{\eeqas}{\end{eqnarray*}}
       
       \newcommand{\bnab}{\mbox{\boldmath ${\nabla}$}}

       \newcommand{\Pc}{ {\mathcal{P}} }

       \newcommand{\be}{{\mathbf e}}
       \newcommand{\bj}{{\mathbf j}}

       \newcommand{\bv}{{\mathbf v}}

       \newcommand{\bB}{{\mathbf B}}

       \newcommand{\bE}{{\mathbf E}}

       \newcommand{\bI}{{\mathbf I}}

       \newcommand{\bOm}{\mbox{\boldmath $\Omega$}}
       \newcommand{\dadb}[2]{\frac{{  d}#1}{{  d}#2}}

\newcommand{\tW}{\tilde{W}}
\documentclass[12pt]{article}
\usepackage[dvips]{graphicx}\usepackage[dvips]{graphics}
\usepackage{enumerate}
\usepackage{psfrag}
\usepackage{color}
\newcommand{\pars}{\partial}
\begin{document}
\begin{center}
{\large \bf On non existence of tokamak equilibria with \\
\vspace{1mm}
purely poloidal flow}
 %On existence of tokamak equilibria with purely poloidal flow}

\vspace{3mm}

{\large  G. N. Throumoulopoulos}

{\it
University of Ioannina, Association Euratom - Hellenic Republic,\\
%\vspace{-1mm}
 Section of Theoretical Physics, GR 451 10 Ioannina, Greece}
\vspace{2mm}

{\large H. Weitzner}

{\it
New York University, Courant Institute of Mathematical Sciences, \\
 %New York,
New York 10012 } \vspace{2mm}

 {\large H. Tasso}

 {\it  Max-Planck-Institut f\"{u}r Plasmaphysik, Euratom
Association,\\
%\vspace{-1mm}
 D-85748 Garching, Germany }
\end{center}
%\noindent
%
%
%\newpage
%\vspace{2mm}
%\begin{center}
%{\large \it December 2000}
%\end{center}
\vspace{2mm}
\begin{center}
{\bf Abstract}
\end{center}

\noindent It is proved that irrespective of compressibility
tokamak steady states with purely poloidal mass flow can not exist
in the framework of either magnetohydrodynamics (MHD) or Hall MHD
models. Non-existence persists  within single fluid plasma models
with pressure anisotropy and incompressible flows.
\vspace{2cm}

%\noindent {\sf Published in  Physics of Plasmas}
\newpage

\begin{center}
{\bf \large I.\  Introduction}
\end{center}

Motivation of the present study was a proof that ideal
magnetohydrodynamics (MHD) steady states of magnetically confined
plasmas with purely poloidal incompressible flows and magnetic
fields having toroidal and poloidal  components can not exist
\cite{Ta,TaTh}. The inconsistency relates to the toroidicity because
in cylindrical geometry equilibria with purely poloidal (azimuthal)
flows exist without restriction on the direction of the magnetic
field \cite{ThTa}. The aforementioned statement of non-existence
includes tokamak equilibria;  only field reversed configurations
with purely poloidal incompressible flows which by definition have
only poloidal magnetic fields are possible. Also, magnetic dipolar
steady states with purely poloidal compressible flows parallel to
the magnetic field were studied in Ref. \cite{KrSo}. In tokamaks,
however, poloidal sheared flows play an important role in the
transition from low to high confinement mode (L-H transition). In
fact, in many experiments
%, including the pioneering
no toroidal rotation in connection with the transition is reported
(see for example Ref. \cite{GrBu}).
 It is
therefore interesting to theoretically examine wether alternative or
additional physical input to incompressibility can remove the
incompatibility.
% of tokamak steady states with purely poloidal flows.

The present study aims at examining whether (i) compressibility (ii)
two fluid effects and (iii) pressure anisotropy can give rise to the
existence of tokamak steady states with purely poloidal flows.  It
turns out that in cases (i) and (ii) as well as  in case (iii) for
incompressible flows the non-existence statement keeps hold.
%We consider either adiabatic
%plasmas with entropy uniform on magnetic surfaces  or magnetic
%surfaces with uniform temperatures which is more appropriate to high
%temperature plasmas.
  It should be noted that there is a number of papers  on equilibria with
 flow  within the framework of MHD \cite{TaTh}-\cite{KrSo},
\cite{ZeGr}-\cite{ThMc},
 the two fluid model \cite{MoSo,McTh,ThMc}, \cite{St1}-\cite{IsHa} and for anisotropic pressure
 \cite{IaBo}-\cite{BeFr}.
 Certain of the  derivations and equations therein are related to the present work.
 We will prefer, however, to present
 the study in a self contained way because explicit  reference to the content of the aforementioned
 papers  for the three models employed would make   a common notation difficult
 and the manuscript inconveniently readable.  Compressible  axisymmetric MHD equilibria is the subject of Sec. II.
 In Sec. III two fluid effects are examined in the
 framework of Hall MHD model with electron temperatures uniform on
 magnetic surfaces and either ion temperatures uniform thereon or
 incompressible ion flows. Pressure anisotropy for incompressible
 flows is considered in Sec. IV.  The conclusions are summarized in
 Sec. V.
 \newpage

 \begin{center}
 {\bf \large II.\ \ Compressibility}
 \end{center}

 The ideal MHD equilibrium states of a  magnetically confined plasma
 are
 governed in standard notation and convenient
 units by the following set of equations:
 %for convinience in Gaussian units with both $4\pi$ and the
 %velocity of light being set to unity: equations}}
 \begin{equation}
 {\bnab} \cdot (\rho {\bv}) = 0,
                         \label{1}
 \end{equation}
 \begin{equation}
 \rho ({\bv} \cdot {\bnab})  {\bv} = {\bj} \times {\bB} -
 {\bnab} P,
                        \label{2}
 \end{equation}
 \begin{equation}
 {\bnab} \times  {\bE} = 0,
                         \label{3}
 \end{equation}
 \begin{equation}
 {\bnab}\times {\bB} = {\bj },
                         \label{4}
 \end{equation}
 \begin{equation}
 {\bnab} \cdot {\bB} = 0,
                        \label{5}
 \end{equation}
 \begin{equation}
 {\bE} +{\bv} \times {\bB} = 0.
                        \label{6}
 \end{equation}
 %$$ h =\frac{M}{ e} $$
 %\beq \mbox{An energy equation equation or equation of state}
                        \label{6a}
 %\eeq
An  energy equation or equation of state needed to close the set of
Eqs. (\ref{1})-(\ref{6}) is not adopted from the beginning; it will
be specified when necessary later. On account of axisymmetry and
Ampere's law (\ref{4}) the divergence-free fields, i.e. the
 magnetic field $\bf B$, the current density $\bf j$, and the
 momentum of the ion fluid element
  $\rho \bv$ can be expressed in terms of  scalar functions
 $\psi(R,z)$, $I(R,z)$, $F(R,z)$ and $\Theta(R,z)$ as
 \beq
 \bB=I(R,z)\bnab \phi + \bnab \phi \times \bnab \psi(R,z),
                              \label{7}
 \eeq
 \beq
 \bj= \Delta^\star\psi\bnab \phi - \bnab \phi \times \bnab
 I(R,z),
                               \label{8}
 \eeq
 \beq
 \rho\bv= \Theta(R,z) \bnab \phi + \bnab \phi
 \times \bnab F(R,z).
                              \label{9}
 \eeq
 Here, $(R,z,\phi)$ are cylindrical coordinates with $z$
 corresponding to  the axis of symmetry;  the functions $\psi$ and
 $F$ label the  magnetic and velocity  surfaces, respectively;
 $\Delta^\star=R^2\bnab\cdot(\bnab/R^2)$.   Expressing the electric field
in terms of the electrostatic potential,  $\bE=-\bnab \Phi$, the
component of Ohms law along $\nabla \phi$ implies that function $F$
is uniform on magnetic surfaces:
 \beq F=F(\psi).
                           \label{9a}
 \eeq
  The electrostatic potential is also a surface quantity
  \beq
  \Phi=\Phi(\psi),
                           \label{10}
 \eeq
  as it follows by projecting (\ref{6}) along
 $\bB$. Two other integrals are identified in terms of surface
 quantities by the components of (\ref{6}) along $\bnab \psi$,
 and (\ref{2}) along $\bB$, respectively:
% They respectively are put in the forms
  \beq
   \frac{1}{\rho R^2}\left(IF^\prime-\Theta\right)=\Phi^\prime,
                             \label{11}
 \eeq
  \beq
   I\left(1-\frac{(F^\prime)^2}{\rho}\right) + R^2 F^\prime
   \Phi^\prime\equiv X(\psi).
                             \label{12}
  \eeq
   Eqs. (\ref{11}) and (\ref{12}) can be solved for the functions
 $I$ and $\Theta$ associated with the toroidal components of $\bB$
 and $\bv$ to yield
 \beq
  I(\psi,R)= \frac{X-R^2\Phi^\prime}{1-M_p^2},
                              \label{13}
  \eeq
  \beq
  \Theta(\psi,R)= \frac{XF^\prime -R^2\rho \Phi^\prime}{1-M_p^2}.
                              \label{14}
  \eeq
   Here $M_p^2$ is the square of the Mach function of the poloidal velocity with respect to the
  Alfv\'en velocity:
  $$
  M_p^2=\left(\frac{v_p}{v_{pA}}\right)^2=\frac{(F^\prime)^2}{\rho}.
  $$
   In
  the present study we will consider both compressible and incompressible flows.
   For typical  tokamak poloidal velocities and temperatures  ($\max v_p\approx
  10^4\
   \mbox{m/sec}$ and $\max kT_i\approx 10\  \mbox keV$) the
   incompressibility condition
   \beq
   \left|\frac{v_p}{v_{th_i}}\right|\ll 1,
                         \label{14a}
   \eeq
   where $v_{th_i}$ is the ion thermal velocity,
   is satisfied and
   therefore incompressible flows are of relevance.  Also, it may
   be
   noted that
    incompressibility is a good approximation for small flows lying within the first elliptic
        regime of the equilibrium differential equations. In MHD this regime is
    \beq
    0<M_p<\beta
                         \label{14b}
    \eeq
        (see for example Ref. \cite{Ha}), where  the maximum value of $\beta$,
        defined as the ratio of the plasma pressure
        to the total (plasma and magnetic field) pressure, is about 0.35 (taking into
        account
        the particularly high values of $\beta$ which have been obtained
        in spherical tokamaks). Note that (\ref{14b}) sets nearly the same bounds on $v_p$
        as (\ref{14a})
        as can be seen by using the definitions of $M_p$ and $\beta$.
       % Also,    the
       % two fluid  equilibrium equations are elliptic when the poloidal ion flow
       % velocity is less than the sound velocity \cite{IsHa}. In tokamak experiments the
%maximum toroidal velocities are on the order of the sound velocity
%and it generally  holds the scaling $v_p\approx 0.1 v_\phi$. Note
%that the transitions from elliptic to hyperbolic regimes are related
%exclusively to $v_p$ because $v_\phi$ is inherently incompressible
%because of axisymmetry. Up to the authors knowledge, such
%transitions potentially  causing confinement degradation because of
%drastic change of the magnetic topology have not been observed in
%tokamak experiments.

  On the basis of (\ref{13}) and (\ref{14}) let us first recover the
  non-existence
  of equilibria with purely poloidal incompressible
  flows and magnetic fields having poloidal and toroidal components.
  On account of incompressibility,
  $\bnab \cdot \bv =0$, (\ref{1}) implies that $\rho=\rho(\psi)$. Then, for
  $\Theta=0$,  (\ref{14})  can not be satisfied, because in addition to surface
  quantities it contains $R$ explicitly, unless either $F^\prime=\Phi^\prime=0$ or $X=\Phi^\prime=0$.
  The former case corresponds to a static equilibrium  ($\bv=0$). The latter  implies that
  $\bv$ is parallel to $\bB$ and therefore $B_\phi=0$ as it also follows from (\ref{13}).
  Therefore, tokamak equilibria with purely poloidal incompressible flows
  are not possible.
  Note that if, inversely, the magnetic field is purely poloidal ($I=0$) inspection of Eqs. (\ref{13})
  and (\ref{14}) implies that  the flow is either purely poloidal or purely toroidal. Therefore
  in field reversed configurations coexistence of toroidal and poloidal incompressible
  flows is not admitted.

  In the case of compressibility the density can vary on magnetic surfaces. For purely poloidal
  flows, however, Eqs. (\ref{13}) and (\ref{14}) imply $I=X(\psi)$ and restrict $\rho$ to be of the form
  \beq
   \rho=\frac{XF^\prime}{\Phi^\prime R^2}.
                            \label{15}
  \eeq
  Evaluation of $\bnab \rho$ on magnetic axis, on which $\bnab \psi=0$, by  (\ref{15}) yields
  \beq
   \bnab \rho_0=-2R_0^3\frac{X_0F^\prime_0}{\Phi^\prime_0}{\be}_R.
                        \label{16}
   \eeq
Therefore,
 \beq
  \bnab \rho_0\neq 0,
                                \label{16a}
 \eeq
 unless $X_0=0$ or $F^\prime_0=0$ which would imply $\rho_0=0$ by
(\ref{15}).
  To proceed further we need an energy equation or equation of state.
  Since  thermal conductivity along $\bB$ is very large in high temperature
 plasmas,  isothermal magnetic surfaces is an appropriate equation of state for tokamaks.
 If $T=T(\psi)$, using the ideal gas law $P=\alpha \rho T$ the
 component of (\ref{2}) along $\bB$  yields for $\Theta=0$
 \beq
 {\bf B}\cdot \bnab\left(\frac{v^2}{2}+ \alpha T \ln \rho\right)=0.
                                                 \label{17}
 \eeq
 or
 \beq
 \frac{v^2}{2} + \alpha T \ln \rho \equiv H(\psi).
                                               \label{18}
 \eeq
  Evaluation of $\bnab \rho$ on axis from (\ref{18}) after multiplying it by
 $\rho$ and using (\ref{9}) for $\bv$, $\Theta=0$ and (\ref{15}) for $\rho$ yields
 \beq
 \left\lbrack H_0-\alpha T_0(1+ \ln \rho_0)\right\rbrack \bnab \rho_0=0.
                                              \label{19}
  \eeq
 If $H_0=\alpha T_0(1+\ln \rho_0)$
   it follows from (\ref{18}) that  $T_0=0$.
 Therefore,  (\ref{19})
  implies $\bnab \rho_0=0$ which contradicts (\ref{16a}).
 As can be shown along the same lines the  non unique definition of $\bnab \rho$
 on axis persists  if alternative equations of state
 are adopted such as isentropic
 magnetic surfaces or barotropic plasmas [$P=P(\rho)$].
 Thus, we can conclude that the non existence of tokamak equilibria with purely poloidal flows
 is extended to the compressible regime.

   The above local proof of non existence can be extended   near the
 magnetic axis.  Indeed,  on account of (\ref{15}),
 Eq. (\ref{18})
  takes the form
  \beq
  \frac{(RF^\prime)^2}{2G^2}\left\lbrack \left(\frac{\pars \psi}{\pars R}\right)^2
  + \left(\frac{\pars \psi}{\pars z}\right)^2 \right\rbrack + \alpha T
  \ln\left(\frac{G}{R^2}\right)
  =H(\psi),
                                          \label{19a}
  \eeq
  where $G(\psi)\equiv XF^\prime/\Phi^\prime$. The $R$-derivative
  of (\ref{19a}) on axis yields
 \beq
  \frac{(R_0F^\prime_0)^2}{G^2_0} \left\lbrack\ \left. \frac{\pars \psi}{\pars
  R}\frac{\pars^2\psi}{\pars R^2}\right|_0\right\rbrack=-2\alpha \frac{T_0
  R_0^3}{G_0}.
                                          \label{19a1}
  \eeq
  Note that the term involving $\left.\pars^2\psi/\pars R\pars z\right|_0$ is not
  included in (\ref{19a1}) because this derivative vanishes as it follows
  from a Mercier expansion based consideration of the generalized Grad-Shafranov
  equation [Eq. (\ref{19c}) below] around the magnetic axis.
  Since the RHS of (\ref{19a1}) is finite, $\left.\pars^2 \psi/\pars
  R^2\right|_0$  must tend to infinity.
 Consequently, the validity of the statement  follows on the
  basis of a Mercier expansion of any solution of (\ref{19a}) near the magnetic axis:
  \beq
  \psi(x,y) =\psi_0 + \frac{1}{2}\left.\frac{\pars^2 \psi}{\pars R^2}\right|_{z_0,R_0} x^2+
     \frac{1}{2}\left.\frac{\pars^2 \psi}{\pars z^2}\right|_{z_0,R_0}
     y^2 + \ldots
                                           \label{19b}
  \eeq
  Here, $(x,y)$ are  Cartesian coordinates defined by $R=R_0+x$ and
  $z=z_0+y$ and $\psi_0$ refers to the magnetic axis.
  Also, the current density on axis becomes singular. Furthermore, it may be noted
   that projection of (\ref{2})
 onto $\bnab \psi$ yields the generalized Grad-Shafranov equation
 \beq
 \bnab\cdot\left\lbrack\left(1-M_p^2\right)\frac{\bnab
 \psi}{R^2}\right\rbrack +\frac{F^\prime
 F^{\prime\prime}}{\rho}\frac{\left|\bnab \psi\right|^2}{R^2}+ \frac{IX^\prime}{R^2} + \rho H^\prime
 +a\rho\left(1-\ln \rho\right)T^\prime=0.
                                    \label{19c}
 \eeq
 Therefore, $\psi$ should satisfy the two differential equations (\ref{17}) and
(\ref{19c})
  containing the  surface quantities $T(\psi)$, $H(\psi)$,  $F(\psi)$ and
$X(\psi)$. As will be shown in the next section for incompressible
flows the solution of these  equations is irrelevant to tokamaks.

 \begin{center}
 {\bf \large III.\ \ Two fluid effects}
 \end{center}

  We will employ the  Hall MHD,
   a simple two-fluid model in the
approximation of very small electron mass. Consequently, the
electron momentum equation  can be  put in the Ohm's law form
  \beq
 {\bE} +{\bv} \times {\bB} = \frac{h}{\rho}\left(\bj \times
 \bB- \nabla P_e\right)
                                              \label{20},
 \eeq
  where $\bv$ is the ion fluid element velocity, $\rho=nM$ and $h\equiv M/e$.
  The right hand side (RHS) of (\ref{20}) contains  the Hall  and
   electron pressure gradient terms. Eq. (\ref{20}) replaces
 (\ref{6}) which is formally recovered in the limit of $h
 \rightarrow \ 0$. The other equations of the model are identical with
 (\ref{1}), (\ref{2}-\ref{5}) on the understanding that
 $\bv=\bv_i$. The momentum equation (\ref{2}) is derived by a superposition
of the electron and ion momentum equations neglecting the electron
 convective velocity term because of the very small electron mass.
 As in Sec. II certain integrals will be   identified in the form of
conserved quantities  on
 magnetic surfaces by projecting (\ref{6}) and (\ref{2})  onto $\bnab \phi$, $\bB$ and  $\bnab \psi$.
 The axisymmetric representations (\ref{7})-(\ref{9}) of the
 divergence free fields will be employed in the derivations to
 follow which
 up to Eq. (\ref{25}) below hold for velocities of
  arbitrary direction.
  Assuming that the electron temperature is uniform on magnetic surfaces
  and using ${\bE}=-\bnab \Phi$
 and   $P_e=\alpha \rho T_e(\psi)$ the component
 of (\ref{6}) along $\bB$ yields
 \beq
 \bB\cdot\bnab \left(\Phi -h\alpha T_e \log \rho \right)=0
                                    \label{21}
 \eeq
 or
 \beq
 \Phi -h\alpha T_e \log \rho \equiv \Xi(\psi).
                                     \label{22}
 \eeq
  %Note that unlike MHD $\Phi$ is not uniform on magnetic surfaces.
 Eq. (\ref{20}) then becomes
 \beq
 \bv \times \bB = \frac{h}{\rho} \bj \times \bB +
 \dadb{\Xi}{\psi}\bnab \psi - h \alpha \dadb{T_e}{\psi}\left(1-\log
 \rho\right)\bnab\psi.
                                     \label{23}
 \eeq
 Projecting (\ref{12}) along  $\bnab \phi$
 yields another integral:
 \beq
 F + h I \equiv f(\psi).
                                     \label{24}
 \eeq
 Eq. (\ref{24}) implies  that in general  the velocity surfaces depart from magnetic
 surfaces. The fact that $\bv$ and $\bB$ share the same surfaces
 in MHD [Eq. (\ref{9a})] is recovered for $h\rightarrow 0$. Furthermore, the component of (\ref{23})
 along $\bnab \psi$ leads to the elliptic differential equation
 \beq
 h \Delta^\star \psi = \rho R^2 \left\lbrack
 \dadb{\Xi}{\psi}-h\alpha\dadb{T_e}{\psi}\left(1-\log
 \rho\right)\right\rbrack + \Theta-\dadb{f}{\psi} I.
                                       \label{25}
 \eeq
  For  $h \rightarrow 0$ (\ref{25}) reduces to the algebraic MHD equation
  (\ref{11}).
  Purely poloidal ion flows will be further considered either
  incompressible or compressible on an individual basis as
follows.\\

 \noindent {\em Incompressible ion flow }  \\

  It is convenient to introduce the
 generalized vorticity
 \beq
 \bOm \equiv \bB + h \bnab \times \bv.
                                                  \label{h16}
 \eeq
 For  purely poloidal flows the $\bOm$-surfaces   coincide with the magnetic surfaces.
 Since  $\bOm$ is divergence free   it can be
 expressed by
 \beq
 \bOm= N(R,z)\bnab \phi + \bnab \phi \times \bnab \psi.
                                               \label{h16a}
 \eeq
Using   (\ref{h16}), $P=P_e + P_i$, $\bE=-\bnab \Phi$ and the
identity
 $$
 \left(\bv\cdot \bnab \right) \bv =\bnab
 \frac{v^2}{2}-\bv\times\bnab\times\bv,
 $$
 Eq. (\ref{2}) can be cast  in the  form
 \beq
 \rho \bnab \tW= \rho \bv \times \bOm - h \bnab P_i,
                                          \label{h30a0}
 \eeq
 where
 \beq
 \tW\equiv h \frac{v^2}{2} + \Phi.
                                          \label{h30a}
 \eeq
 The component of (\ref{h30a0}) along $\bnab \phi$ yields
 \beq
 F=F(\psi).
                                           \label{h31}
 \eeq
  Eq. (\ref{h31}) implies that $\bv$ lies on magnetic
 surfaces and therefore
  $
   \rho=\rho(\psi)
  $
 because of incompressibility.
 Note that the electron fluid element velocity lies on
 magnetic surfaces too whatever is the direction of $\bv_e$ as it follows from the electron momentum equation
 with the $(\bv_e\cdot\bnab)\bv_e$ term being neglected.
 From (\ref{22}) and (\ref{24}) then it follows $\Phi=\Phi(\psi)$
  and $I=I(\psi)$.
  Also, substituting  the expressions for $\bB$ and $\bv$ from  (\ref{7}) and
 (\ref{9}) into (\ref{h16a}) leads to the following expression for $N$:
 \beq
 N=I + h \bnab\left(\frac{F^\prime}{\rho}\right)\cdot \bnab \psi +
 h \frac{F^\prime}{\rho}\Delta^\star \psi.
                                       \label{h22}
 \eeq
 Projecting (\ref{h30a0}) onto $\bOm$ and $\bnab\psi$, respectively,
 and using (\ref{h22}) for  $N$ furnishes
 \beq
 h P_i = h P{_{is}}(\psi) - \rho \tW,
                                           \label{h32}
 \eeq
\beq
 hM_p^2(\psi)\Delta^\star\psi+\frac{1}{2}\dadb{M_p^2}{\psi} \left|\bnab \psi\right|^2
 =R^2\left(h\dadb{P_{is}}{\psi} - \dadb{\rho}{\psi}\Phi(\psi)\right)
-I(\psi)\dadb{F}{\psi}.
                                                              \label{h72}
 \eeq
 Here, the surface quantity $P_{is}(\psi)$ for vanishing flow coincides with the ion
 pressure and $M_p^2=(F^\prime)^2/\rho$.  Eqs. (\ref{h72}) and
(\ref{25}) for $\Theta=0$ can be cast in the forms
 \beq
  \Delta^\star\psi=-f(\psi)-R^2g(\psi),
                                \label{h51}
  \eeq
  \beq
  \left|\bnab\psi\right|^2=2\left\lbrack i(\psi)
+R^2j(\psi)\right\rbrack,
                               \label{h52}
  \eeq
  where
  $f$, $g$, $i$ and $j$ are known functions of
  $\rho$, $F^\prime$, $T_{e}$, $P_{is}$ $\Phi$ and $I$. The forms of Eqs. (\ref{h51}) and (\ref{h52})
  indicate that
  the magnetic surfaces are identical in shape  with those of the Palumbo
  solution \cite{Pa} which, however, can not describe tokamak equilibria
  ($\psi$ contours of this solution are provided in Figures 1 and 2 of Ref. \cite{BiTa}).
  Note that the pressure is not
  uniform on magnetic surfaces and therefore the equilibrium is not isodynamic.  Only
  under the additional assumption $P=P(\psi)$, (\ref{h32}) implies
 $B^2=B^2(\psi)$. Recapitulating, two fluid effects in the frame
 of Hall-MHD model result in non pertinent to tokamaks isodynamic like equilibria with purely poloidal
 incompressible flows.\\

 \noindent {\em Compressible ion flow }  \\

 In connection with tokamak equilibria we will assume that additionally to the electron temperature the ion
 temperature is uniform on magnetic surfaces. Some of the
 derivations in the previous part of this section, i.e. Eqs. (\ref{22}), (\ref{24}), (\ref{25}) and
(\ref{h30a0}), remain valid. Also, for purely poloidal flows
 irrespective of compressibility the
 velocity surfaces coincide with the magnetic surfaces
[$F=F(\psi)$]
 and therefore $I=I(\psi)$ by (\ref{24}). Using $P_i=\alpha\rho
 T_i(\psi)$ the component of (\ref{h30a0}) along $\bB$ yields
 \beq
  \tW+h\alpha T_i \ln \rho \equiv \Lambda(\psi).
                                          \label{29}
  \eeq
  %where $\Lambda(\psi)$ is a ``constant of integration" surface
  %quantity.
    From (\ref{22}), (\ref{h30a}) and (\ref{29}) eliminating the functions $\Phi$ and $\tW$
  we obtain for $\Theta=0$
  \beq
   h\frac{(F^\prime)^2}{\rho^2}\frac{|\bnab\psi|^2}{R^2} =
  \Lambda(\psi)-\Xi(\psi)-h\alpha (T_e+T_i)\ln\rho.
                                           \label{30}
  \eeq
  Acting the gradient operator on (\ref{30}) and evaluating the resulting equation
  on   magnetic axis leads to
  $$\frac{T_{i0}+T_{e0}}{\rho_0}\bnab \rho_0=0$$ and therefore
  $\bnab \rho_0 = 0$.
  On the other side, on account of (\ref{24}) and $I=I(\psi)$, (\ref{25}) for
   $\Theta=0$ becomes
   \beq
   h \Delta^\star \psi = \rho R^2 \left\lbrack
  \Xi^\prime -h\alpha T_e^\prime \left(1-\log
 \rho\right)\right\rbrack -\left(F^\prime+hI^\prime\right) I.
                                       \label{31}
  \eeq
   Evaluating the gradient of (\ref{31}) on magnetic axis leads to
 \beq
  R_0^2 \left(\Xi^\prime_0+h\alpha
 T_{e0}^\prime\ln\rho_0\right)\bnab\rho_0=-2R_0\rho_0\Xi^\prime_0\be_{R}
 + h\left\lbrack\bnab\Delta\psi^\star_0 + 2\alpha
 R_0T_{e0}^\prime\rho_0\left(1-\ln \rho_0\right)\be_R\right\rbrack.
                                    \label{31a}
 \eeq
  A prerequisite that (\ref{31a}) is compatible with $\bnab\rho_0=0$
  is that the RHS of this relation vanishes.  The RHS of (\ref{31a}), however, consists of
  the first large
   MHD-like  term
  and the  second   term on the order of $h$ which for tokamaks is small. Therefore,
  these terms  can not cancel each other. Note that the first term involving the product
  $\rho_0 \Xi_0^\prime$ remains always finite because if
   $\Xi^\prime_0$ becomes very small,  this in the MHD-like limit of $h\rightarrow0$
  would imply  very large densities by (\ref{15}) which keeps valid in this
  limit.
   Therefore, it follows  from (\ref{31a})
  that $\bnab \rho_0\neq 0$. It may be noted that the assumption of  weak enough two-fluid equilibrium effects
  leading to non-unique definition of $\bnab \rho_0$ on axis is fulfilled   in tokamaks  particularly in the
  major interior part of the plasma.
  %It should be  clarified that the above proof of  non-unique definition of $\bnab \rho_0$ on axis presumes
  % week enough two-fluid equilibrium effects   which is a typical situation particularly in the
  %major internal part
  %of tokamak plasmas.
  The inconsistency  in connection with  $\bnab \rho_0$
   persists if alternative ion equations of state
  are adopted such as barotropic [$P_i=P_i(\rho)$] or isentropic ion velocity  surfaces.
  Therefore, as in MHD, tokamak equilibria with purely poloidal
  compressible flows in the framework of Hall-MHD model can not
  exist.
  % {\bf As in the case of MHD } the proof is local.

  \begin{center}
 {\bf \large IV.\ \ Pressure anisotropy}
 \end{center}

 We will employ an one fluid model with the momentum equation (\ref{2})
 replaced by
 \begin{equation}
 \rho ({\bv} \cdot {\bnab})  {\bv} = {\bj} \times {\bB} -
 {\bnab}\cdot \Pc,
                        \label{32}
 \end{equation}
 where the tensor
 \beq
 \Pc=P_\parallel \bI + \sigma \bB\bB
                              \label{33}
  \eeq
 is associated with the pressure anisotropy, a measure of which
 is the  quantity
 \beq
  \sigma\equiv \frac{P_\parallel-P_\perp}{B^2}.
                                   \label{33a}
 \eeq
  The other  equations (\ref{1}) and (\ref{3})-(\ref{6}) remain
  unchanged. Double adiabatic equilibria can be obtained in the
  framework of the Chew Goldberger Low equations of state \cite{ChGo}:
  $P_\parallel=S_\parallel(\psi)\rho^3/B^2$ and $P_\perp=S_\perp(\psi)\rho
  B$ where $S_\parallel$ and $S_\perp$ are arbitrary surface quantities.
  Also, adiabatic MHD equilibria with isotropic pressure can be
  recovered by setting $P_\parallel=P_\perp=S(\psi)\rho^\gamma$,
  where $S$ is the specific entropy. Here we will consider
  incompressible flows without other specification of $P_\parallel$
  and $P_\perp$.
  The integrals (\ref{9a})-(\ref{11})
  stemming from the Ohm's law remain valid. The component of (\ref{32}) along
  $\bB$ yields
  \beq
  \bB\cdot\bnab\left\lbrack I\left(1-M_p^2-\sigma\right)
  +R^2F^\prime \Phi^\prime \right\rbrack=0
                                           \label{33b}
  \eeq
  or
  \beq
  I\left(1-\frac{(F^\prime)^2}{\rho}-\sigma\right)+R^2F^\prime\Phi^\prime\equiv
  X(\psi).
                                                   \label{34}
   \eeq
  For isotropic pressure [($\sigma=0$)],  (\ref{34}) reduces to
 (\ref{12}). As in the isotropic case (\ref{11}) and (\ref{34}) can
 be solved for $I$ and $\Theta$ to yield
  \beq
  I= \frac{X-R^2\Phi^\prime}{1-M_p^2-\sigma},
                              \label{35}
  \eeq
  \beq
  \Theta= \frac{XF^\prime -R^2\rho
  \Phi^\prime(1-\sigma)}{1-M_p^2-\sigma}.
                              \label{36}
  \eeq
  For purely poloidal  flows (\ref{35}) and (\ref{36}) imply
  \beq
  \rho=\frac{XF^\prime}{R^2 \Phi^\prime(1-\sigma)}
                              \label{37}
  \eeq
  and
  \beq
  I=\frac{X}{1-\sigma}.
                        \label{38}
  \eeq

  If $\sigma=\sigma(\psi)$ it follows from (\ref{37})  that equilibria with
  purely poloidal incompressible flows are not possible (when $B_\phi \neq 0$ and $B_p\neq
  0$) because then these flows  are not compatible with $\rho=\rho(\psi)$ .
  {\em Static }  equilibria with $\sigma=\sigma(\psi)$
  were investigated  in Refs. \cite{MeKo,Cle}. Compatibility of
  (\ref{37}) with $\rho=\rho(\psi)$
   restricts $\sigma$ to be of the form
   \beq
   \sigma= 1-\frac{f(\psi)}{R^2},
                             \label{39}
  \eeq
  where $f(\psi)$ is an arbitrary surface quantity.
  Eq. (\ref{38}) then implies  $I=R^2 X(\psi)/f(\psi)$. Apparently,
  possible equilibria of this kind are irrelevant to tokamaks in
  which the toroidal magnetic field, $B_\phi=I/R$, must have a vacuum
  component proportional to $R^{-1}$.

\begin{center}
{\large \bf V.\ \ Conclusions}
\end{center}

We have shown that the non-existence of ideal MHD tokamak equilibria
with purely poloidal incompressible flow  can be extended to the
cases of (i) compressible MHD flows (ii)  Hall-MHD incompressible
and compressible flows
 %steady states regardless compressibility
 and
(iii) one fluid equilibria with pressure anisotropy and
incompressible flows. The non-existence relates to the toroidicity.
Specifically, for incompressible flows an inconsistency appears in
terms of relations which in addition to surface quantities have an
explicit dependence on $R$
%on the radial distance  $R$ from the axis of symmetry
or only isodynamic-like equilibria are possible and for
compressible flows the density gradient does not have unique
definition on the magnetic axis.  Also, for MHD compressible flows
the proof  can be extended near the magnetic axis.

The existence of this kind of equilibria on account of additional
physical input, e.g. in plasmas with pressure anisotropy and
compressible flows,  or other models  remains an open question.
Experimental clarification on whether an even small toroidal
velocity is necessary for the L-H transition to occur would be
helpful. If not necessary, the existence of tokamak equilibria with
purely poloidal flows from the theoretical point of view should
remain an even more challenging problem.
\newpage
\begin{center}
 {\large\bf Acknowledgements }
 \end{center}

 The authors would like to thank the anonymous referee for critical comments,
 and particularly for his statement on  an  extension of
 the local proof  which  resulted in an
 improved version of the work.

 Part of this work was conducted during a visit of  the author G.N.T.
 to the Max-Planck-Institut f\"{u}r Plasmaphysik,
 Garching. The hospitality of that Institute is greatly appreciated.

 The present work was performed under the Contract of Association
ERB 5005 CT 99 0100 between the European Atomic Energy Community
and the Hellenic Republic. The views and opinions expressed herein
do not necessarily reflect those of the European Commission.

 \end{document}